\documentstyle[prd,aps,preprint,tighten]{revtex}

\begin{document}
\draft
\title{Evidence for mass dependent effects in the spin
structure of baryons}

\author{Johan Linde\footnote{E-mail: jl@theophys.kth.se} and
H{\aa}kan Snellman\footnote{E-mail: snell@theophys.kth.se}}
\address{Department of Theoretical Physics\\
Royal Institute of Technology\\
S-100 44 STOCKHOLM\\
SWEDEN}

\maketitle
\begin{abstract}
We analyze  the axial-vector form factors of the
nucleon hyperon system in a model with mass dependent quark spin
polarizations.
This mass dependence is deduced from an earlier analysis\cite{jlhs,jlhs2} of
 magnetic moment data, and implies that the spin contributions from the
 quarks to a baryon decrease with the mass of the baryon. When applied to
the axial-vector form factors, these mass dependent spin polarizations
bring the various sum-rules from the
model in better agreement with experimental data. Our analysis
 leads to a reduced value for the total spin
polarization of the proton.
\end{abstract}


\section{Introduction}
In two earlier papers\cite{jlhs,jlhs2}
we have discussed the magnetic moments
of baryons in an
extension of the quark model, which allows for general flavor symmetry
breaking and where the quark magnetic moments are allowed
to vary with the isomultiplet $B$. The magnetic moments of the baryons in
this model can be written as a linear sum of contributions from the various
flavors
 \begin{equation}
        \mu(B^{i}) = \mu_{u}^{B}\Delta u^{B^{i}} +
        \mu_{d}^{B}\Delta d^{B^{i}} + \mu_{s}^{B}\Delta
        s^{B^{i}},
        \label{moment}
\end{equation}
where $\mu_{f}^{B}$ is an effective magnetic moment of the quark of flavor
$f$ in the isomultiplet $B$ and $\Delta f^{B^{i}}$
is the corresponding spin polarization for baryon $B^{i}$, $i$ being the
baryon charge state. By symmetry arguments the $\Delta f^{B^{i}}$'s in the
octet baryons can be expressed as constant linear
combinations of the three $\Delta f$'s for the proton, which are the only
spin polarizations needed to describe the octet:
\begin{equation}
        \Delta f^{B^{i}} = \sum_{f'}M(B^{i})_{ff'}\Delta f',
        \label{deltabi}
\end{equation}
where $f,f'$ runs over $u,d,s$, and the $M(B^{i})$'s are matrices with
constant elements. In particular for the six mirror symmetric
baryons of type $B(xyy)$, where $x$ and $y$ are different flavors, we have
$\Delta y^{B^{i}}=\Delta u$, $\Delta x^{B^{i}}=\Delta d$ and $\Delta
z^{B^{i}}=\Delta s$ where the flavor $z$ is the non-valence quark flavor.
In the non-relativistic quark model (NQM)
the values of these spin polarizations are
$\Delta u = \frac{4}{3}$, $\Delta
d = -\frac{1}{3}$ and $\Delta s = 0.$

Due to the homogeneity of the
right hand side of (\ref{moment}), it is a question
of definition if the dependence on the baryon
multiplet is considered to be associated with
the quark magnetic moment rather than with
the spin polarization. In Refs.\cite{jlhs} and \cite{jlhs2} we have chosen to
analyze the data by keeping the spin polarizations fixed throughout.

Here we will analyze the opposite situation, where the
spin polarization is instead assumed to vary with the baryon multiplet and the
quark magnetic moments are the same for all multiplets.

This scheme has the
advantage of making the properties of the quarks static and environment
independent. Since the effective magnetic moment of a quark in the NQM has the
form
\begin{equation}
        \mu_{f}=\frac{e_{f}}{2m_{f}},
\end{equation}
$e_{f}$ being the quark charge, this means that there is no dependence of the
effective quark mass $m_{f}$ on $B$. This is in accordance with
the fact that the same constituent
quark masses can be used successfully to predict the baryon octet and
decuplet masses with only a hyperfine splitting interaction in the
Hamiltonian.
The disadvantage is that the spin structure varies from multiplet to
multiplet.

The most important further merit of this interpretation, and the one that
we are going to analyze here, is that the sum-rules governing
the axial-vector form factors are better fulfilled in this scheme, although
the errors are still somewhat large to definitely decide between
either of the two ways of attributing the mass dependence effect.

\section{Calculating the model parameters}
The spin structure parameters in the expressions for the
magnetic moments and in the
deep inelastic scattering experiments and axial-vector
form factors are not a priori the
same. In many models they are nevertheless
proportional\cite{karl}, and can be normalized to be the same. We
normalize them to the axial-vector form factor $g_A^{np}=1.2573$, as is
generally done.

We write the baryon magnetic moments as
\begin{equation}
                \mu(B^{i}) = \sum_{f,f'}\mu_{f}\alpha(B)
M(B^{i})_{ff'}\Delta f' ,
\end{equation}
where $\mu_{f}$ is the magnetic moment of the
quark of flavor $f$ and $\Delta f$
is the corresponding spin polarization. The factor $\alpha(B)$ is an overall
factor, the same for all flavors, depending only on the isomultiplet $B$. The
flavor symmetry breaking is then accounted for by the quark magnetic
moments that are free parameters. This symmetry breaking is assumed to be
the same for all isomultiplets.

In our previous analysis we associated the factor $\alpha(B)$ with the quark
magnetic moments and defined $\mu_f^{B}=\alpha(B)\mu_f$ as in equation
(\ref{moment}). We can choose to normalize $\alpha(B)$ to $\alpha(N)=1$, in
which case $\mu_{f}^{N}=\mu_{f}$. The other values
of $\alpha$ can then be obtained from the previously
extracted values of the $\mu_{d}^{B}$'s
as\cite{jlhs,jlhs2}
\begin{eqnarray}
    \alpha(\Lambda) & = & \mu_{d}^{\Lambda}/\mu_{d}^{N} = 0.88\pm
    0.04,\label{alphaL}\\
        \alpha(\Sigma) & = & \mu_{d}^{\Sigma}/\mu_{d}^{N} = 0.91\pm 0.01,  \\
        \alpha(\Xi) & = & \mu_{d}^{\Xi}/\mu_{d}^{N} = 0.85\pm
0.03.\label{alphaX}
\end{eqnarray}
We will now instead associate $\alpha(B)$ with the spin polarizations.
Equation (\ref{deltabi}) is then rewritten as
\begin{equation}
        \Delta f^{B^{i}} = \sum_{f'}M(B^{i})_{ff'}\alpha(B)\Delta f'.
        \label{nydeltabi}
\end{equation}
Thus, {\it e.g.\/} in the mirror symmetric baryons $B(xyy)$, we instead
have $\Delta y^{B^{i}}=\alpha(B)\Delta u$, $\Delta
x^{B^{i}}=\alpha(B)\Delta d$ and $\Delta z^{B^{i}}=\alpha(B)\Delta s$.

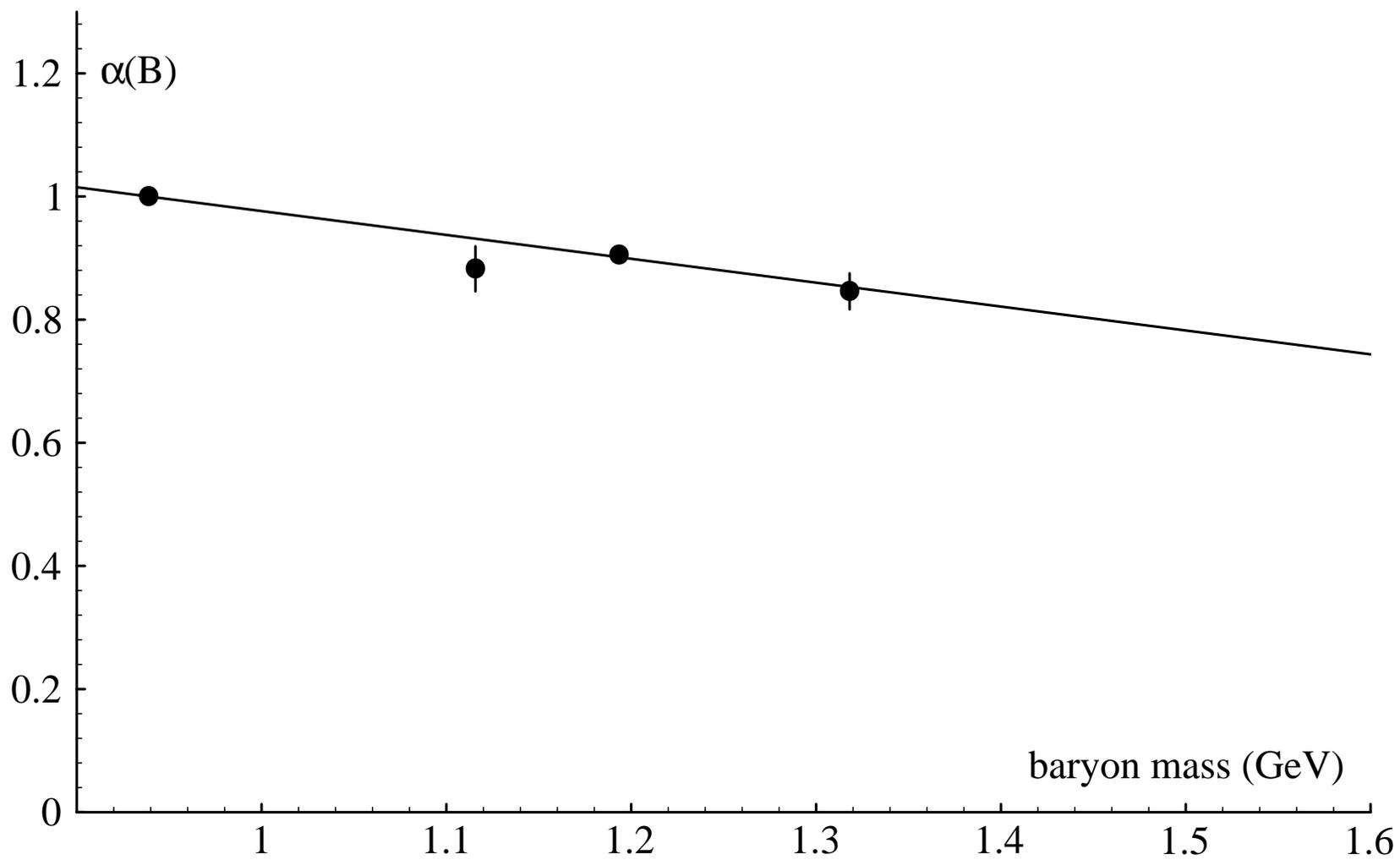
\begin{figure}
\caption{The ratio $\alpha(B)=\mu_d^B/\mu_d^N$ as a function of the baryon
mass. The points are the data for the nucleon, $\Lambda$, $\Sigma$ and $\Xi$
as given by equations (\protect\ref{alphaL})-(\protect\ref{alphaX}).
The straight line represents the linear fit according
to equation (\protect\ref{lin}).}
\label{cool_plot}
\end{figure}

 The values of $\alpha(B)$
can be well fitted to a linear function of the mean mass of  $B$ as
shown in Fig.~\ref{cool_plot}. The linear relation is
\begin{equation}
        \alpha(m)= 1-0.376(m - 0.939),
        \label{lin}
\end{equation}
when $m$ is expressed in GeV. We will continue to use
$\alpha(B)\equiv\alpha(m_B)$ in the following.

If this relation is extrapolated to the decuplet resonances,
it can be tested by
measuring some of their magnetic moments. It is
then possible to fit the expression for $\mu(\Omega^{-})$ to obtain the
value of $\alpha(\Omega^{-})$\cite{jlhs2}.

The most remarkable property of this fit is that the
quark spin polarization of a
baryon, and thus also the contribution form its quark magnetic moment,
vanishes at $m \approx 3.6$ GeV, provided the linear relation does not
break down before we reach this value.

We will nevertheless test this linear relation in
what follows by using the interpolated $\alpha$'s from the equation above.
These values are
\begin{eqnarray}
    \alpha(\Lambda) & = & 0.93\pm 0.02, \\
        \alpha(\Sigma) & = & 0.90\pm 0.02 , \\
        \alpha(\Xi) & = & 0.86\pm 0.02.
\end{eqnarray}
To illustrate why this $B$ dependent factor is needed we regard
the sum-rule
\begin{equation}
        \mu(p) +\mu(\Xi^{0}) +\mu(\Sigma^{-}) - \mu(n) - \mu(\Xi^{-}) -
        \mu(\Sigma^{+}) = 0,
\end{equation}
which follows when the quark magnetic moments and spin polarizations both
are independent of $B$. It  is badly broken by the
experimental data so that the
left hand side is instead $0.49\pm 0.05$.
In our more general parameterization this sum-rule becomes
\begin{equation}
         \mu(p) +\frac{\mu(\Xi^{0})}{\alpha(\Xi)} +
         \frac{\mu(\Sigma^{-})}{\alpha(\Sigma)}
         - \mu(n) - \frac{\mu(\Xi^{-})}{\alpha(\Xi)} -
         \frac{\mu(\Sigma^{+})}{\alpha(\Sigma)} = 0.
 \end{equation}
Due to the construction of the $\alpha$'s this sum-rule is satisfied.

\section{The axial-vector form factors}
Whether we associate the $\alpha(B)$ factors to the quark magnetic moments
or the spin polarizations, obviously does not affect
the analysis of the magnetic moments. However, the
analysis of the axial-vector form factors will be modified when we let the
spin polarizations be given by (\ref{nydeltabi}).

The axial-vector form factors can in this parameterization now be written
\begin{eqnarray}
        g_{A}^{np}        & = & \Delta u - \Delta d , \\
        g_{A}^{\Lambda p} & = & \frac{1}{3}(2\Delta u - \Delta d - \Delta
        s)\alpha(\Lambda) , \\
        g_{A}^{\Xi \Lambda} & = & \frac{1}{3}(\Delta u + \Delta d - 2\Delta
        s)\alpha(\Xi)  ,\\
        g_{A}^{\Sigma n} & = &(\Delta d - \Delta s)\alpha(\Sigma) .
\end{eqnarray}
This can be used to derive the two sum-rules
\begin{eqnarray}
                \frac{g_{A}^{\Xi \Lambda}}{\alpha(\Xi)} +
                 \frac{g_{A}^{\Lambda p}}{\alpha(\Lambda)} & = &
                \frac{g_{A}^{\Sigma n}}{\alpha(\Sigma)} + g_{A}^{np},  \\
        \frac{g_{A}^{\Xi \Lambda}}{\alpha(\Xi)} +
        g_{A}^{np} & = & 2 \frac{g_{A}^{\Lambda p}}{\alpha(\Lambda)},
\end{eqnarray}
which are barely satisfied without the $\alpha$'s.
The relations are satisfied as follows
\begin{eqnarray}
(0.98\pm 0.07)\quad     1.07\pm 0.06 & = & 1.04\pm 0.09 \quad (1.06 \pm
0.08), \\
(1.51\pm 0.05)\quad     1.55\pm 0.06 & = & 1.57\pm 0.05 \quad (1.46 \pm 0.03),
\end{eqnarray}
corresponding to the two equations above. The numbers in parentheses are
the values without the $\alpha$'s (i.e. $\alpha \equiv 1$).
The experimental values
$g_A^{np}=1.2573\pm 0.0028$ and $g_A^{\Xi\Lambda}=0.25\pm0.05$ are taken
from the Review of Particle Properties data table\cite{data},  the
value $g_A^{\Sigma n}=-0.20\pm0.08$ from Hsueh {\it et al.\/}\cite{hsueh}
and the value $g_A^{\Lambda p}=0.731\pm0.016$ from Dworkin {\it et
al.\/}\cite{dworkin}. For a more detailed discussion we refer the reader to
Sec. 3 of Ref.\cite{jlhs}.

Although the improvement relative
to the case without the $\alpha$'s  might not be
dramatic, both sum-rules are definitely better satisfied with the $\alpha$'s.

As a further test we calculate the constant $R=\frac{\Delta u-\Delta
d}{\Delta u-\Delta s}$ defined in Ref.\cite{jlhs}. This
constant has the value $R=1.18\pm0.01$ from the magnetic moment data. Our
expression for this constant, expressed in terms of axial-vector
form factors, is now
\begin{equation}
        R=\frac{2g_{A}^{np}}{ g_{A}^{\Lambda p}/\alpha(\Lambda) + g_{A}^{\Xi
        \Lambda}/\alpha(\Xi)+
        g_{A}^{\Sigma n}/\alpha(\Sigma)+g_{A}^{np}} = 1.19\pm0.06.
\end{equation}
 This is again an improvement over the value $R=1.23\pm0.06$
 found in Ref.\cite{jlhs}.

The four axial-vector form factors can
be parameterized by two variables, which
 we choose as  $\Delta u- \Delta d=g_A^{np}$
 and $a_8=\Delta u +\Delta d-2\Delta s$.
 Another often used parameterization is $F+D=\Delta u-\Delta d$,
 $F/D=\frac{\Delta u-\Delta s}{\Delta u+\Delta s-2\Delta d}$.

 Since $g_A^{np}=1.2573\pm 0.0028$ is
 by far the best measured parameter we will use
 this as a fix parameter and express the
 three other axial-vector form factors in terms of
 $g_A^{np}$ and $a_8$. This
 gives
 \begin{eqnarray}
 \frac{ g_A^{\Lambda p}}{\alpha(\Lambda)} & = &
        \frac{1}{6}a_8 + \frac{1}{2}g_A^{np}, \\
 \frac{ g_A^{\Sigma n}}{ \alpha (\Sigma)}& = & \frac{1}{2}a_8
 -\frac{1}{2}g_A^{np} , \\
        \frac{g_A^{\Xi\Lambda}}{\alpha(\Xi)}& = & \frac{1}{3}a_8.
 \end{eqnarray}

 We have performed two least square fits of $a_8$ using these formulas and
 the experimental numbers quoted above, one with the $\alpha$'s and one
 without. With the $\alpha$'s we get $a_8=0.89\pm0.08$ with $\chi^{2}=0.31$
 and without the $\alpha$'s we get $a_8=0.70\pm 0.08$ with $\chi^{2}=1.9$.
 We see that there is a considerable improvement when the $\alpha$'s are
 included. These values correspond to $F/D=0.75\pm 0.04$ and $F/D=0.63 \pm
 0.04$ respectively.

We find a rather large deviation from
 the value  $F/D=0.575 \pm 0.016$ found by Close and Roberts\cite{close1},
 who deliberately have chosen not
 to include the induced form factor $g_2$ and use the value
  $g_A^{\Sigma n}=-0.340 \pm 0.017$\cite{data}. The deviation is also
  rather large relative to the
 value $a_8=0.601\pm 0.038$ used by Ellis and Karliner\cite{Ellis} in
their proton spin polarization analysis. The somewhat drastic increase in
 error comes from the unfortunately rather
 poor determination of the induced form factor
 $g_2$.

 Finally we remark that our evaluation of the isospin symmetry breaking
 parameter $T=\mu_u/\mu_d$ from the spin polarization\cite{jlhs} of
 the nucleon is not affected by this reinterpretation, since it is based on
 the ratio $\Delta\Sigma /g_{A}^{np}$ which
 is the same in both interpretations.

 \section{Implications for the proton spin polarization analysis}
 A change in the value of $a_8$ has a non-negligible
 influence on the proton spin
 polarization analysis. We will illustrate this using the formulas  from
the analysis of Ellis
 and Karliner\cite{Ellis}. Their evaluation of
 $\Delta\Sigma=0.31\pm 0.07$ can be
 expressed as
 \begin{equation}
        \Delta\Sigma(Q^2) = 9\frac{\Gamma_1^p(Q^2)
        -(\frac{g_A^{np}}{12}+\frac{a_8}{36})f(\alpha_s)}{h(\alpha_s)},
 \end{equation}
 where
 \begin{equation}
        f(\alpha_s)=1-\left(\frac{\alpha_s(Q^2)}{\pi}\right)-
        3.5833\left(\frac{\alpha_s(Q^2)}{\pi}\right)^2
        -20.2153\left(\frac{\alpha_s(Q^2)}{\pi}\right)^3-
        {\cal O}(130)\left(\frac{\alpha_s(Q^2)}{\pi}\right)^4
 \end{equation}
 and
 \begin{equation}
        h(\alpha_s) =1-\left(\frac{\alpha_s(Q^2)}{\pi}\right)-
        1.0959\left(\frac{\alpha_s(Q^2)}{\pi}\right)^2-
        {\cal O}(6)\left(\frac{\alpha_s(Q^2)}{\pi}\right)^3.
 \end{equation}
 The constant $g_A^{np}$ has the usual value $g_A^{np}=1.2573$,
 but the constant $a_8$ has
 in Ref.\cite{Ellis} the value $a_8=0.601\pm 0.038$.

 The value of $\Delta\Sigma$ will change appreciably if we
 change the value of $a_8$. Let the change in $a_8$ be denoted
 $\delta a_8$, and the new value of $\Delta\Sigma$ be denoted
 $\Delta\Sigma'$. We then have
 \begin{equation}
        \Delta\Sigma'=\Delta\Sigma-\frac{\delta
        a_8}{4}\left(1-{\cal O}(2)
        \left(\frac{\alpha_s(Q^2)}{\pi}\right)^2 \right)\approx
        \Delta\Sigma-\frac{\delta a_8}{4}.
 \end{equation}
 The value of $a_8 =0.89 \pm 0.08$ found above  will
 thus lead to a different estimate of
 the total spin polarization of the proton.
 $\Delta\Sigma$ will change to
 \begin{equation}
        \Delta\Sigma'=0.31- \frac{0.29}{4}=0.24\pm0.09.
 \end{equation}
 In our previous analysis of isospin symmetry breaking in the
 baryon magnetic moments\cite{jlhs} this value
 favors a slightly smaller isospin symmetry
 breaking than the value $\Delta\Sigma=0.31$. It also changes slightly the
 quark spin content of the proton to the values
 \begin{eqnarray}
        \Delta u & = & 0.86\pm 0.04 , \\
        \Delta d & = & -0.40\pm 0.04 ,\\
        \Delta s & = & -0.22\pm 0.05.
 \end{eqnarray}
 calculated by means of $\Delta\Sigma'$, $a_8$ and $g_A^{np}$. The main
 effect is to allow $\Delta s$ to be larger. This is consistent with the
 magnetic moment analysis in Ref.\cite{jlhs}.

 \section{Discussion and conclusions}
 As we have seen above there is supportive evidence from the axial-vector
 form factor data that the spin polarizations of the quarks are diminishing
 with the increase of mass of the host particle. This mass dependence is
 born out in the sum-rules that can be written and are well satisfied by
 the experimental data.

 One possible interpretation of this effect could be that the quarks simply
 lose their orientation as the excitation energy increases, and finally,
 at $3.6$ GeV, become totally unoriented on the time
 scales considered here, much
 as if they were enclosed in a heat bath.

 As the total angular momentum of the
 proton is fixed to $1/2$, this means
 that there must be a contribution from some other
 electrically neutral component that
 increases its angular momentum with baryon mass to
 compensate for the decrease in the
 contribution coming from the quarks.

 One possibility is to attribute such a
 contribution to the presence of gluonic
 components in the baryons.
 This is perfectly consistent with the findings from deep
 inelastic scattering experiments, that only about
 half of the proton momentum is
 carried by the quarks.  Also a collective mode of the Skyrmion type, with
a rather small contribution to the magnetic moment, could be envisaged to
 manifest in this way.

 The new feature found here is that this contribution
 varies linearly with the mass of the baryon multiplet.

 This could {\it e.g.\/} be the case if to this
 contribution there is associated a moment of
 inertia that grows with mass, but has stationary angular velocity.

 As an {\it ad hoc} example we can consider
 a proton spin sum-rule of the form
 \begin{equation}
        \frac{1}{2} = \frac{1}{2} \Delta\Sigma + \frac{1}{2} I \cdot {\rm
const},
 \end{equation}
 where $I$ is the moment of inertia of the non-quark component of the
 nucleon. When $I \propto m $, $m$ being the
 baryon mass, we can rewrite this as
 \begin{equation}
        \Delta\Sigma(m) =\Delta\Sigma(m_p) - (m-m_p)\cdot c'  ,
 \end{equation}
 which leads to equation (\ref{lin}).

 The puzzling outcome of this is that it predicts the quark spin polarization,
 and thus also the magnetic moment of a baryon, to vanish at $m \approx
 3.6$ GeV, provided the linear relation between $\alpha$ and $m$ is
 still valid there.

 All this emphasizes the
 importance of trying to measure the magnetic moments of high mass baryon
 states and also to try to calculate them with lattice gauge techniques.

 The studies performed by Leinweber {\it et al.\/}\cite{draper}
 supports indirectly the findings
 here. Their lattice gauge calculations have been done in quenched QCD
 by keeping the spin polarization fixed to the NQM values. The quark
 magnetic moments then show a decrease in value with increasing mass of the
 host particle, in much the same way as we found in Ref.\cite{jlhs2}, where we
 also chose the keep the spin polarizations mass independent.

 Also lattice gauge calculations of the axial-vector form factors for the
 heavier states, would possibly shed light on the behavior found here. Such
 calculations have already been performed for the nucleon system\cite{axial}.

 Finally we have shown how a change in the evaluation of the axial-vector
 coupling constants will affect the proton spin polarization analysis. Our
 value for the constant $a_8$ favors a lower value of the proton quark
 spin sum.

 \acknowledgements
This work was supported
by the Swedish Natural Science Research
Council (NFR), contract F-AA/FU03281-308.

\end{document}